\documentclass[preprint2]{aastex631}
\usepackage{amsmath}
\usepackage{graphicx}
\usepackage{txfonts}
\usepackage{orcidlink}

\begin{document}

\title{Ionized envelopes around protoplanets and the role of radiative feedback in gas accretion}

\correspondingauthor{Mat\'ias Montesinos}
\email{matias.montesinosa@usm.cl}

\author[0000-0001-9789-5098]{Mat\'ias Montesinos}
\affiliation{Departamento de Física, Universidad Técnica Federico Santa María, Avenida España 1680, Valparaíso, Chile}
\affiliation{Millennium Nucleus for Planet Formation, NPF, Valparaíso, 2340000, Chile}

\author[0000-0002-7056-3226]{Juan Garrido-Deutelmoser}
\affiliation{Instituto de Astrof\'isica, Pontificia Universidad Cat\'olica de Chile, Santiago, Chile}
\affiliation{Millennium Nucleus for Planet Formation, NPF, Valparaíso, 2340000, Chile}

\author[0000-0003-1965-3346]{Jorge Cuadra}
\affiliation{Departamento de Ciencias, Facultad de Artes Liberales, Universidad Adolfo Ibáñez, Av.\ Padre Hurtado 750, Viña del Mar, Chile}
\affiliation{Millennium Nucleus for Planet Formation, NPF, Valparaíso, 2340000, Chile}

\author[0000-0002-8065-4199]{Mario Sucerquia}
\affiliation{Univ. Grenoble Alpes, CNRS, IPAG, 38000 Grenoble, France}

\author[0000-0003-3713-8073]{Nicolás Cuello}
\affiliation{Univ. Grenoble Alpes, CNRS, IPAG, 38000 Grenoble, France}

\author[0000-0003-3903-8009]{Matthias R. Schreiber}
\affiliation{Departamento de Física, Universidad Técnica Federico Santa María, Avenida España 1680, Valparaíso, Chile}
\affiliation{Millennium Nucleus for Planet Formation, NPF, Valparaíso, 2340000, Chile}

\author[0000-0003-1385-0373]{Mar\'ia Paula Ronco}
\affiliation{Instituto de Astrofísica de La Plata, CCT La Plata-CONICET-UNLP, Paseo del Bosque S/N (1900), La Plata, Argentina}
\affiliation{Millennium Nucleus for Planet Formation, NPF, Valparaíso, 2340000, Chile}

\author[0000-0001-8577-9532]{Octavio M. Guilera}
\affiliation{Instituto de Astrofísica de La Plata, CCT La Plata-CONICET-UNLP, Paseo del Bosque S/N (1900), La Plata, Argentina}
\affiliation{Millennium Nucleus for Planet Formation, NPF, Valparaíso, 2340000, Chile}



\begin{abstract}
Planetary growth within protoplanetary disks involves accreting material from their surroundings, yet the underlying mechanisms and physical conditions of the accreting gas remain debated. This study aims to investigate the dynamics and thermodynamic properties of accreting gas giants, and to characterize the envelope that forms near the planet during accretion. We employ three-dimensional hydrodynamical simulations of a Jupiter-mass planet embedded in a viscous gaseous disk. Our models incorporate a non-isothermal energy equation to compute gas and radiation energy diffusion and include radiative feedback from the planet. Results indicate that gas accretion occurs supersonically towards the planet, forming an ionized envelope that extends from the planetary surface up to 0.2 times the Hill radius in the no-feedback model, and up to 0.4 times the Hill radius in the feedback model. The envelope’s radius, or ionization radius, acts as a boundary halting supersonic gas inflow and is pivotal for estimating accretion rates and H$\alpha$ emission luminosities. Including radiative feedback increases accretion rates, especially within the ionization radius and from areas to the right of the planet when the star is positioned to the left. The accretion luminosities calculated at the ionization radius are substantially lower than those calculated at the Hill radius, highlighting potential misinterpretations in the non-detection of H$\alpha$ emissions as indicators of ongoing planet formation.
\end{abstract}



\section{Introduction} \label{sec:intro}

Planets grow by accretion (e.g., \citealt{Pollack+1996}), and during their formation process, it is thought that a gaseous envelope should surround them. The most common scenario states that a circumplanetary disk (CPD) forms around gas giants during the last stages of their formation while the circumstellar disk is still present. Protoplanets candidates are often seen as hot spots in disks around young stars, where the temperature and density are high enough for them to be detected, giving insight into the early stages of planet formation. The distinct thermal signatures of these hot spots are typically observed peaking at L'-band (e.g., \citealt{Reggiani+2014}), and have been shown to be associated with the accretion luminosity \citep{Montesinos+2015}. The planet-disk interaction structures can be observed using the new generation of adaptive optics (e.g., \citealt{Benisty+2022, Currie+2022}), which allows for studying the physical and chemical properties of protoplanetary disks. Directly observing planet-forming processes within protoplanetary disks remains a significant challenge in astrophysics. Currently, only two CPDs have been confirmed using adaptive optics imaging, both within the protoplanetary disk system PDS 70 \citep{Keppler+2018, Isella+2019, Muller+2018, Christiaens+2019}. In addition to these confirmations, several CPD candidates have been found (e.g.,
\citealt{Reggiani+2018, Tsukagoshi+2019, Boccaletti+2020, Pinte+2020}).

The structure of CPDs and the accreting gas has been first studied through two-dimensional numerical experiments of steady-state flows (e.g., \citealt{Miki1982, Korycansky+1996}) followed by more performant simulations including viscous heating and radiative effects (e.g., \citealt{Lubow+1999, D'Angelo-2003, Ward&Canup2010}). However, two-dimensional hydrodynamical simulations to describe the gas accretion in the surroundings of a CPD cannot fully describe the nature of the accreting flow, especially because the scale height of a CPD around a Jupiter mass planet should be of the order of the Hill sphere. Newer numerical simulations assuming \textit{isothermal} disks in three dimensions were able to follow the gaseous flow in more detail, showing that the gas accretes from high altitudes nearly through the vertical direction towards the planet (e.g., \citealt{Machida+2008, Ayliffe+2009, Szulagyi+2014}).

Despite such numerical progress, there is a discussion on how to characterize the radial and vertical extent of CPDs and the flux of matter falling into the planet's surface. \cite{Ayliffe+2009} suggest that the angular momentum distribution of the circumstellar disk could determine the edges of CPDs or the planet envelope. At the edges, a peak in the angular momentum of the gas is noticeable, indicating the radial position where the gas is bound to the CPD rather than the circumstellar disk. Other criteria propose that the rim of a CPD is located where the tidal torque from the star removes angular momentum of particles located at the edges defining a truncation radius  \citep{Martin+2011}. Also, three-dimensional simulations showed a significant increment of the azimuthally averaged radial velocity of the gas, indicating a location where particles are escaping from the Hill sphere \citep{Tanigawa+2012}, defining thus the radius of the CPD. In all these models, the envelope radius is about $\sim 0.2-0.5$ times the planet's Hill radius. 

A key ingredient to describe the gaseous flow around protoplanets is the inclusion of a \textit{non-isothermal} energy equation describing heating and cooling processes.   By implementing radiative transfer calculations, new models showed that the gas temperature in the surroundings of a protoplanet is crucial in describing the accretion flow and the formation of CPDs or circumplanetary envelops (e.g., \citealt{Szulagyi+2016, Lambrechts+2019}). For instance, under some circumstances, a circumplanetary envelope develops within the Roche lobe when the gas close to the planet is hot enough. Furthermore, because the gas falls vertically towards the planet, a shock front is also expected to arise, liberating a hot and luminous shock front in the envelope.

Since the gaseous envelope of accreting planets is a viscous fluid, the conversion of mechanical energy into radiation through viscous dissipation is possible. Assuming that the total energy of the falling material is converted into radiation, the accretion luminosity is given by $L_{\rm acc} \simeq G M_{\rm p} \dot{M}_{\rm p}/R_{\rm p}$, where$R_{\rm p}$ should correspond to the radius of the planet and $\dot{M}_{\rm p}$ to the planet's accretion rate. However, the precise mechanism governing accretion onto the planet remains a subject of debate. For instance, it is unclear at which radius the mass flow can be best characterized  or how the energy content of the envelope affects the accretion rate. To address the latter point, non-isothermal models are necessary. Therefore, the computations of accretion and luminosity must be revisited.

The production of shocks during the accretion process can result in the emission of H$\alpha$ (\citealt{Szulagy+Mordasini+2017}). Such emissions are expected to arise in the vicinity of accreting planets during the vertical accretion flow phase, which occurs at nearly free-fall velocities \citep{Aoyama2018, Aoyama2019}. Today, H$\alpha$ emission is the most commonly utilized tracer for protoplanetary accretion  (e.g.,\citealt{Close+2014, Zhu2015, Zurlo+2020, Szulagyi+2020, Huelamo+2022}). Computing $L_{H\alpha}$ is challenging; moreover, if the physics of the accreted gas is unknown. However, a reasonable approach leads to an almost direct relation between planet accretion and $L_{H\alpha}$, i.e., $L_{H\alpha} \propto \dot{M}_{\rm p}$ (\citealt{Aoyama2019}). Despite these approximations, several key parameters remain unidentified.
For instance, the geometry of the surface shocks and their ability to convert mechanical energy into radiation energy remains unclear. Furthermore, H$\alpha$ production also depends on the ionization level of the surrounding gas, which has been poorly studied in the context of protoplanetary disks. Also, it is not clear at which radius $\dot{M}_{\rm p}$ should be computed to correctly characterize $L_{H\alpha}$.  The unclear factors may explain the absence of H$\alpha$ emissions from accreting planets in observations \citep{Zurlo+2020}, despite theoretical predictions of higher accretion rates (e.g., \citealp{Aoyama2019}).

In the present work, we follow the evolution of a protoplanetary disk with an embedded Jupiter mass planet focusing in the vicinity of the planet near the Hill sphere. Since we are interested in the early stages of planet formation, the Jupiter mass planet is still accreting material into its surface, liberating thus internal energy associated with its formation process. Therefore, we assume that the planet has an intrinsic radiative feedback component consisting of a constant luminosity locally injected by the planet into its surroundings at a constant rate.

Our modeling considers non-isothermal thermodynamics in which we include a time-dependent energy equation to treat the transport of energy of the gas and the radiative energy of photons via a coupled equation. This treatment allows us to obtain a realistic temperature field and the radiative flux diffusion of energy. From the obtained temperature and density field, we compute the thermal ionization degree of the gas, assumed to be pure Hydrogen gas. We identified a hot partially ionized envelope surrounding the planet, which is smaller than the Hill sphere and can be characterized by a specific \textit{ionization radius}. Such radius defined from pure thermodynamic arguments) is consistent with previous estimations of the bound limits of CPDs or circumplanetary envelops. We compute the accretion rate at this radius and independently the envelope luminosity released within this frontier. As expected, the envelope luminosity matches the accretion luminosity. However, the accretion rate depends on the energy content of the envelope and, therefore, on the radiative feedback from the planet. 


\section{Accretion onto circumplanetary disks and radiative feedback}\label{Accretion}

\subsection{Planetary feedback}\label{planetfeedback}

We assume that the planet radiates away all its gravitational potential energy, which is parametrized as follows
\begin{equation}\label{Lfeedback}
    L_{\rm p} = \frac{G M_{\rm p}}{R_{\rm p}} \dot{M}_{\rm pebble} =  \frac{G M_{\rm p}}{R_{\rm p}} \frac{M_{\rm p}}{\tau},
\end{equation}
the parameter $\tau = M_{\rm p}/\dot{M}_{\rm pebble}$ corresponds to the planetary mass doubling time-scale, where $M_{\rm p}$ is the planetary mass, and $\dot{M}_{\rm pebble}$ is the pebble accretion rate. The physical radius of the planet is given by $R_{\rm p} = (3 M_{\rm p} / (4 \pi \rho_{\rm solid}))^{ (1/3)} $, where we use $\rho_{\rm solid} = 3 ~ {\rm g\,cm^{-3}}$ as the mean density of the planet. The energy released by the planet per unit time, $L_{\rm p}$, is injected at a constant rate during the simulations.

It is worth mentioning that $\dot{M}_{\rm pebble}$  is a free parameter characterized by the doubling time $\tau$. This pebble accretion rate is assumed responsible for the intrinsic planet luminosity (radiative feedback) $L_{\rm p}$ (Equation \ref{Lfeedback}). In principle, the pebble accretion rate should be a function of time (e.g., \citealt{Garate+2021}). However, for simplicity, we will assume it to be constant, as in \cite{Montesinos+2015} and \cite{Benitez-Llambay+2015}. Also, $\dot{M}_{\rm pebble}$ must not be confused with the gas accretion onto the planetary envelope. We will discuss the planetary accretion rate in the next section.

\subsection{Planet accretion}\label{PlanetAcretion}

We compute the net flux mass passing through the envelope by directly measuring the mass of the envelope and differentiating it numerically between snapshots. In spherical coordinates, centered at the location of the planet, the volume of the envelope is characterized by an accretion radius ($r_{\rm acc}$), therefore;
\begin{equation}\label{eqMdot}
    \dot{M}(t) = \frac{d}{dt} \oiiint_{r_{\rm acc}} \rho dV, 
\end{equation}
where $\rho$ is the volume gas density, and $dV = r^2 \sin{\theta} d\theta d\phi$ is a volume element. We note that the total volume must be evaluated at $r_{\rm acc}$. 

Typically, the accretion radius $r_{\rm acc}$ can be constrained by two different measurements. The first measures the distance at which the disk's gas with sound speed $c_{\rm s}$ will be gravitationally bound to the planet, such distance is normally called the Bondi radius. This quantity is obtained by equating the escape velocity to the sound speed at the planet location $c_{\infty}$, i.e., $r_{\rm acc} = R_{\rm Bondi} = G M_{\rm p} / c_{\infty}^2$, where $M_{\rm p}$ is the mass of the planet.

The second possibility assumes the planet's Hill radius as the accretion radius, i.e., $r_{\rm acc} = R_{\rm Hill} = r_{\rm p} \left(  M_{\rm p} / (3 M_\star) \right)^{1/3}$, where $M_{\rm p}$ and $M_\star$ are the masses of the planet and star, respectively, and $r_{\rm p}$ the radial distance from the star to the planet. Within this radius, the attraction of the planet dominates over the gravity of the star.

Based on these definitions, it is normally assumed that a gas parcel will be bound to the planet if: $r < R_{\rm Bondi}$ and $r < r_{\rm H}$, i.e., $r_{\rm acc} = \rm{min}(R_{\rm Bondi}, R_{\rm Hill})$. This work will introduce another definition for the accretion radius related to the accretion envelope. Such definition is conditioned by the planet's ionized region characterized by an \textit{ionization radius}, which will be discussed in Section \ref{ionizacion_section}.

\section{Thermal ionization}\label{ionizacion_section}

Assuming a primordial gas composition in thermal equilibrium, the ionization degree of a single atomic species can be described by the Saha equation \citep{Saha1921}. If we neglect external ionization sources such as cosmic rays, stellar X-rays, or UV radiation (e.g., \citealt{Rybicki_Radiative_processes_1979}), we obtain
\begin{equation} \label{saha}
\frac{n_{\rm p} n_{\rm e}}{n_{\rm H}} = 2 \left( \frac{g_{\rm i}}{g_{\rm n}} \right)  \left( \frac{2 \pi m_{\rm p} k T}{h^2} \right)^{3/2} \exp{ \left\{-\frac{\chi}{k T}\right\}},
\end{equation}
where $n_{\rm p}$, $n_{\rm e}$, $n_{\rm H}$ denote the number density of proton, electron, and hydrogen, respectively. The constants $k$, $h$, and $m_{\rm p}$ are the Boltzmann constant, the Planck constant, and the proton mass, respectively. The partition functions $g_{\rm i}$ and $g_{\rm n}$ correspond to the ionized and neutral species, respectively, where we assume $g_{\rm i} \sim g_{\rm n}$, reasonable for low-ionization gases at relatively low temperatures. The temperature of the gas is described by $T$. The factor 2 accounts for the two possible states of an electron. For the ionization potential $\chi$, we assume hydrogen atoms with $\chi = 13.6$ eV.

The total number density is represented by $n = \rho / (\mu m_{\rm p})$. If we assume that the number of electrons, $n_e$, is roughly equal to the number of protons, $n_{\rm p}$, and the hydrogen density is defined as $n_{\rm H} = n - n_{\rm p}$. We can express an ionization fraction, $X$, as the ratio of electrons to the total number density, $X \equiv n_{\rm e} / n$. Thus, the Saha equation (Equation ~\ref{saha}) can be transformed into a second-order equation in terms of the ionization fraction as
\begin{equation}\label{fractionX}
    X^2 + 2 \frac{\Gamma(T)}{n} X - 2 \frac{\Gamma(T)}{n}  = 0,
\end{equation}
where $\Gamma(T)$ is defined by
\begin{equation}\label{gamma}
     \Gamma(T) \equiv \left( \frac{2 \pi m_{\rm p} k T}{h^2}  \right)^{3/2} \exp{ \left\{-\frac{13.6\;{\rm eV}}{k T}\right\}}.
\end{equation}
Solving Equation \ref{fractionX} for $X$, we find
\begin{equation}\label{eqX}
    X = \frac{- 2 \frac{\Gamma(T)}{n}  \pm 2 \sqrt{\frac{\Gamma(T)^2}{n^2} + 2 \frac{ \Gamma(T)}{n}}}{2}.
\end{equation}
The ionization fraction $X$ (Equation \ref{eqX}) gives a measure of the fraction of atoms or ions that are in an ionized state in the gas. It ranges between $0 \leq X \leq 1$, with the factor $1/2$ applied in Equation \ref{eqX} to normalize the results. The gas is considered fully ionized when $X = 1$ and completely neutral when $X = 0$. When $X < 0.5$, the gas is considered to be essentially neutral, whereas, for values of $X \geq 0.5$, the gas is partial to fully ionized. In addition, it is worth noting that $X$ increases with temperature $T$ and decreases with number density $n$. Beyond certain thresholds in both $T$ and $n$, the ionization fraction $X$ approaches an asymptotic value of unity, indicating a fully ionized gas. The case $X=0.5$ then lies very close to this transition regime where the gas changes rapidly from predominantly neutral to nearly fully ionized.

The condition $X = 0.5 $ is expected to be fulfilled within the disk at some specific radius from the planet. Using Equation \ref{eqX}, it is possible to compute the surface surrounding the planet for which this transition occurs. Inside this surface, the gas is partially or fully ionized $X > 0.5$. Beyond it, in the outward direction, the gas becomes neutral $X < 0.5$. We will call this radius the \textit{ionization radius} $R_{\rm ion}$, corresponding to the \textit{effective} or \textit{accretion} radius discussed in Section \ref{PlanetAcretion}. This quantity is key in setting the boundaries when computing the planet accretion rate with Equation~\ref{eqMdot} and the envelope luminosity (see Section \ref{RadTransfSection}) as discussed in the next subsection.

\subsection{The ionization radius and the accreting surface}

The ionization radius, $R_{\rm ion}$, is a key parameter to study planetary accretion rates. It represents the radial distance from the planet at which the degree of ionization of the surrounding gas reaches $X=0.5$. This value defines the boundary of the ionized region in the disk, which in turn determines the location of the accretion surface. The degree of ionization, $X$, computed in Equation \ref{eqX}, is expected to be very low in standard protoplanetary disks (e.g., $X \sim 10^{-12}$, \citealt{ArmitageBook}) but can increase to a significant level ($X \sim 0.5$) in regions surrounding accreting planets, due to the high temperatures reached $T \sim 10^3-10^4$ K as reported by \citealt{Szulagyi+2016}.

Since we aim to study the behavior of the gas within the boundaries of the ionized envelope produced during the accretion process, we need to evaluate the planet accretion rate defined by Equation \ref{eqMdot} at the ionization radius by replacing $r_{\rm acc}$ with $R_{\rm ion}$ in Equation \ref{eqMdot}, therefore we obtain;
\begin{equation}\label{eqMdot_final}
\dot{M}(t) = \frac{d}{dt} \oiiint_{X=0.5} \rho dV,
\end{equation}
where the iso-surface (bounded by $X = 0.5$) contains the volume $V$, and the degree of ionization, $X$, is computed from Equation~\ref{eqX}. The ionization radius, $R_{\rm ion}$, is computed using planetocentric coordinates, giving a family of radial vectors pointing from the planet to the ionization surface defined by $X=0.5$.

\section{Energy equation}\label{RadTransfSection}
\label{lum_section}
We assume a two-temperature solver approach to describe the energy transport, which solves two coupled energy equations. Both equations describe the transport of the gas energy $e_{\rm g}$ and the radiative energy of photons $e_{\rm rad}$. We use a gray approximation under the flux-limited diffusion theory \citep{Levermore+1981}. Adopting the formalism from \cite{Bitsch+2013}, we have:
\begin{multline}\label{energyEq2}
    \frac{\partial e_{\rm rad}}{\partial t} + \nabla \cdot \textbf{F} = \rho \kappa_{\rm P} \left( B(T) - c e_{\rm rad} \right),
\end{multline}

\begin{multline}\label{energyEq1}
    \frac{\partial e_{\rm g}}{\partial t} + (\textbf{u} \cdot \nabla) e_{\rm g} =
    - P \nabla \cdot \textbf{u} - \rho \kappa_{\rm P} \left( B(T) - c e_{\rm rad} \right) \\
    + Q^+_{\rm v} + Q^+_{\rm p} + S,
\end{multline}
where $\textbf{F}$ is the radiative flux, $\rho$ is the gas density, $\kappa_{\rm P}$ the Planck-mean opacity, $c$ the speed of light, $\textbf{u}$ the gas velocity, $P$ the gas pressure, $B(T) = 4 \sigma_{\rm SB} T^4$ the radiative source term given by black-body energy radiation density, being $\sigma_{\rm SB}$ the Stefan-Boltzmann constant, $Q^+_{\rm v}$ the viscous dissipation function, $Q^+_{\rm p}$ the flux of radiative energy received from the planet (feedback), and $S$ the stellar heating which we assume $S = 0$. The gas energy $e_{\rm g}$ is related to the temperature $T$ through $e_{\rm g} = \rho c_{\rm V} T$, where $c_{\rm V}$ is the specific heat at constant volume. The system is closed by using an ideal gas equation given by $P = R_{\rm gas} \rho T / \mu$, where $\mu = 2.3 ~\rm  g ~ mol^{-1}$  is the mean molecular weight valid for a standard solar mixture. The gas energy is also given by $e_{\rm g} = R_{\rm gas} \rho T / \mu (\gamma - 1)$, and therefore, the gas pressure takes the form $P = e_{\rm g} (\gamma - 1)$, where $\gamma = 1.43$ is the adiabatic index.

The radiative flux $\textbf{F}$ is computed using the flux-limited theory \citep{Levermore+1981}, which enables us to compute the radiation energy flux in a regime compatible with both the optically thick and thin limit. Therefore we have:

\begin{equation}\label{Flux1}
    \textbf{F} = \frac{\lambda c }{\rho \kappa_{\rm R}} \nabla e_{\rm rad},
\end{equation}
where $\kappa_{\rm R}$ is the Rosseland mean opacity and $\lambda$ the flux limiter coefficient where we use the approximation by \cite{Kley+1989}.

From energy conservation arguments, the total (bolometric) luminosity is calculated as \( L = \int dV d\epsilon/dt \), where \( \epsilon \) is the total energy density \( \epsilon = e_{\rm rad} + e_{\rm g} \), and \( dV = r^2 \sin{\theta} d\theta d\phi dr \) is a volume element in spherical coordinates centered on the planet. Here, the energies associated with gas and radiation are obtained from Equations \ref{energyEq2} and \ref{energyEq1}.

Since we are interested in the energy content of the envelope and its total released luminosity, we restrict the calculations to regions where \( X \geqslant 0.5 \) (where the gas is partially or fully ionized); therefore, we have:

\begin{equation}\label{eq:Lshock}
    L_{\rm envelope} = \frac{d}{dt} \oiiint_{X=0.5} \epsilon dV,
\end{equation}
where \( X \) is the ionization fraction computed from Equation \ref{eqX}. By differentiating the total energy with respect to time, the resulting luminosity from the evolving envelope is equal to the accretion luminosity of material falling onto the core which includes contraction and release of internal heat (\citealt{Mordasini+2012}).

Also, the accretion luminosity can be computed from,

\begin{equation}\label{eq:Lacc}
    L_{\rm acc} = GM_{\rm core} \dot{M}_{\rm core} \left ( \frac{1}{R_{\rm in}} - \frac{1}{R_{\rm out}} \right ),
\end{equation}
here, $\dot{M}_{\text{core}}$ represents the characteristic core accretion rate, $M_{\text{core}}$ is the mass of the core ($\simeq M_p$), $R_{\text{in}}$ is the closest distance the material reaches to the core, and $R_{\text{out}}$ is the region where core accretion begins.

The accretion luminosity corresponds to the liberated gravitational energy of the envelope surrounding the protoplanets. Thus, it must match the envelope luminosity calculated through equation \ref{eq:Lshock}.

\section{Numerical simulations}\label{numerical}

\begin{table*}[]
\caption{Simulation parameters}
\footnotesize
\centering
\begin{tabular}{c c c c}
\hline 
\hline
Simulation & planet mass $[M_{\rm J}]$ & opacity $[\text{cm}^2/\text{g}]$ & doubling timescale $[yr]$ \& planet luminosity $[L_\odot]$ \\
\hline

 i & 1  &  1 & $\tau = \infty$ ~ $\Rightarrow  L_{\rm p} = 0$  \\
 ii & 1 &  1 & $\tau = 2.0 \times 10^5$ ~ $\Rightarrow L_{\rm p} = 1.5 \times 10^{-3} $  \\
 iii & 1  &  0.01 & $\tau = \infty$ ~ $\Rightarrow  L_{\rm p} = 0$  \\
 iv & 1 &  0.01 & $\tau = 2.0 \times 10^5$ ~ $\Rightarrow L_{\rm p} = 1.5 \times 10^{-3} $  \\
 
\hline
\hline
\end{tabular}
\flushleft
\vspace{0.1cm}
\normalsize
\label{PlanetTable}
\end{table*}

We use the publicly available hydro-code FARGO3D \citep{Benitez-Llambay+2016} to follow the evolution of a protoplanetary disk with an embedded planet. The planet is located at the equatorial plane at the intersection between cell interfaces in azimuth, radius, and co-latitude, lying at the center of an eight-cell cube. Therefore, to compute $Q^+_{\rm p}$ (in Equation \ref{energyEq1}), we assume that the energy released by the planet per unit time per cell (radiative feedback) is simply its luminosity $L_{\rm p}$ (obtained from Equation \ref{Lfeedback}) divided by 8 \citep{Benitez-Llambay+2015}.

The gravitational potential of the planet is given by $\Phi = - G M_p/\sqrt{(r^2 + s^2)}$, where $r$ is the distance to the planet, and $s$ serves as a potential smoothing length calculated by $s = \text{aspect ratio} \times (r/r_p)^{\text{flaring index}} \times r \times \text{Thickness Smoothing}$. This work uses an aspect ratio of 0.05, a flaring index of 0, and a Thickness Smoothing value of 0.1. The planet is located at $r_p = 1$ au and is not allowed to migrate. The choice of $r_p = 1$ AU is motivated by its common use in studies of planet formation and the expectation that gas accretion onto forming giant planets is efficient in this region (e.g., \citealp{Guilera+2020}).

In our modeling, we include a module to solve the radiative transfer equations described in Section \ref{RadTransfSection}.  The non-stationary energy equation (Equations \ref{energyEq2}-\ref{energyEq1}) includes a viscous heating term only without stellar irradiation. For simplicity, we adopt the constant-viscosity prescription with a typical kinematic viscosity $\nu = 7 \times 10^{13} \rm cm^2 s^{-1}$ (or $\nu = 1 \times 10^{-5}$ in code units) to model the disk turbulent viscosity and a constant effective opacity of $\kappa = 1~cm^2~g^{-1}$ (fiducial model). Additionally, we ran two identical models where the only change was reducing the opacity to 0.01 $~cm^2~g^{-1}$.

To ensure that the disk remains optically thick, we use the effective optical depth prescription from \cite{Hubeny1990} (i.e., $\tau_{\text{eff}} = \sqrt{\frac{3}{4} + \frac{3\tau}{8} + \frac{1}{4\tau}}
$), verifying that the chosen opacity values are consistent with the optically thick regime in the regions of interest. This is necessary to ensure that the planet luminosity interacts with the surrounding material, as radiative transport is primarily governed by the optical depth rather than by small changes in opacity.

We use a Minimum Mass Solar Nebula model (e.g., \citealt{Crida2009}) to describe the surface density profile,

\begin{align}
    \Sigma(r) &= 6 \times 10^{-4} \left(\frac{r}{r_{\rm p}}\right)^{-0.5} M_\odot~\rm{au}^{-2}, \\ 
    &= 5.33 \times 10^3 \left(\frac{r}{r_{\rm p}}\right)^{-0.5} \rm{g} \, \rm{cm}^{-2}.
\end{align}

We do not prescribe a gap for the protoplanet; instead, we allow the disk to evolve from the initial profile.

The numerical grid has a resolution of $256 \times 128 \times 64$, corresponding to azimuth $\phi$, radius $r$, and co-latitude $\theta$ cells (respectively) in spherical coordinates. The computational domain spans an azimuthal extent from $-\pi/2$ to $\pi/2$, a radial range from $0.5$ to $1.5 r_{\rm p}$ (with the planet located at $r_{\rm p} = 1$ au), and a colatitudinal span from approximately $\theta_{\rm min} \simeq 7.4^\circ$ (above the midplane) to $\theta_{\rm max} = 90^\circ$ (equatorial plane). The simulation focuses only on the half of the disk above the midplane. The  cell interfaces are evenly spaced along each dimension. We run two simulations characterized by the planet parameters given in Table \ref{PlanetTable}.

\subsection{Ionization radius}\label{ionization_radius}

\begin{figure*}
    \centering
\includegraphics[height=4cm, width=\textwidth]{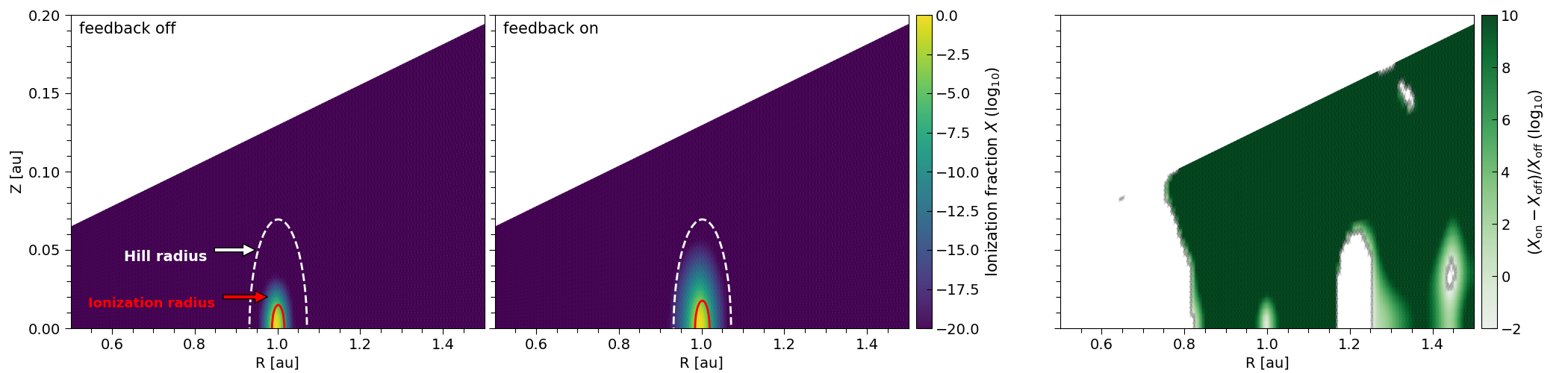}
\caption{Ionization degree $X$ of the disk for models with and without feedback in the final output of the simulation (1000 planetary orbits). A region of ionization, characterized by $X \geq 0.5$, was discerned close to the planet. A specific radius bounds this region (red curve close to the planet), referred to as the \textit{ionization radius}. The ionization radius in the model without feedback (left panel) is approximately 0.34 $R_{\rm Hill}$. Upon feedback activation (middle panel), the ionization radius increased to approximately 0.42 $R_{\rm Hill}$. The rightmost panel shows the effective gain when passing from feedback \textit{on} to \textit{off}, i.e., $(X_{\rm on} - X_{\rm off})/X_{\rm off}$. Blank regions indicate that the ionization level remains unchanged whether the feedback is included or not.} 
\label{fig:saha}  
\end{figure*}

Once the temperature is obtained from the energy equation, we can compute the ionization degree by solving Equation \ref{eqX}. In Figure \ref{fig:saha}, we plot the ionization fraction $X$ of the disk for the last output of the simulation (1000 orbits) comparing two simulations: feedback off/on cases (simulations 1 \& 2, see Table \ref{PlanetTable}).

\begin{figure}
    \centering
    \includegraphics[width=\columnwidth]{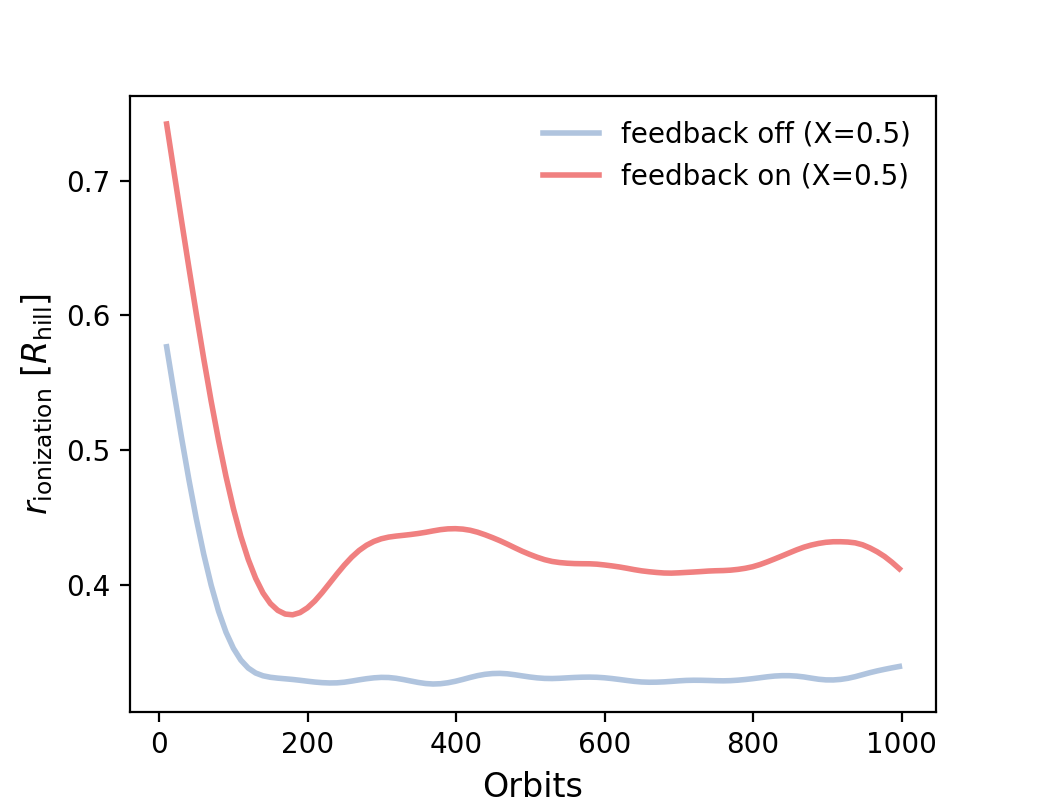} 
\caption{Evolution in time of the ionization radius for models with and without feedback. During the first 200 planetary orbits, the planet carves a cavity, and the ionization radius decreases. After a quasi-stationary regime is reached (at $\sim 300$ orbits), the ionization radius remains relatively constant, with a maximum value of $0.34 R_{\rm Hill}$ (feedback off) and $0.42 R_{\rm Hill}$ (feedback on).}
    \label{fig:Racc}
\end{figure}

In both situations, regions close to the planet are partially or fully ionized ($X>0.5$), indicating that the planet's envelope is hot enough to trigger thermal ionization. As discussed in Section~\ref{ionizacion_section}, the ionized region is limited by a specific ionization radius ($R_{\rm ion}$) pointing from the planet to the extent surface where the condition $X = 0.5$ is reached. The red curve inside Figure \ref{fig:saha} shows the corresponding ionization radius. For the feedback-off model, this radius reaches $\sim 0.34 \, R_{\rm Hill}$, while when the feedback is turned on it reaches $\sim 0.42 \, R_{\rm Hill}$, resulting in an increment of about 24\% when the planet's feedback is activated. In the innermost regions of the disk, situated in proximity to the central star and the midplane, the gas exhibits an ionization level of approximately $X \sim 10^{-10}$ (left panels Figure \ref{fig:saha}), which is typical for the inner regions of protoplanetary disks.

The rightmost panel of Figure \ref{fig:saha} shows the effective gain when the feedback is included, i.e., $(X_{\rm on}-X_{\rm off})/X_{\rm off}$. Interestingly, the feedback modifies the ionization levels of a large gas column from the midplane to the top of the disk (green area in Figure \ref{fig:saha}). However, the ionization level is extremely low $X \sim 10^{-20}$; therefore, a gain factor $\sim 10$ in $X$ still leaves the disk fully neutral everywhere (except inside the ionization radius). The blank regions indicate that the ionization level remains unchanged whether the feedback is included or not. 

Figure \ref{fig:Racc} shows the ionization radius's evolution as a time function. The ionization radius is a family of vectors pointing from the planet's center to a region where the ionization degree reaches $X = 0.5$. In principle, each radius magnitude may be different in different directions. Therefore, we took its maximum magnitude to characterize a unique ionization radius plotted for each simulation snapshot in Figure \ref{fig:Racc}. The blue curve corresponds to a model without feedback, while the red one includes it. Both curves show similar evolutionary trend. At the initiation of the simulation, the ionized region is situated between 60-80\% of the Hill radius, a measure of planetary gravitational influence. As the planet continues to interact with the disk, it carves a cavity, and the disk approaches a state of quasi-stasis, leading to a reduction in the extent of the ionized region and its sphere of influence. In the quasi-steady regime (after 300 planetary orbits), the feedback-off model shows a maximum ionization radius of about $R_{\rm ion}^{\rm max} \sim 0.33 R_{\rm Hill}$. At the same time, for the feedback-on case, it reaches $R_{\rm ion}^{\rm max} \sim 0.43 R_{\rm Hill}$, representing an \textit{enhancement} of about $\sim 28\%$ when the feedback is activated. The maximum ionization radius $R_{\rm ion}^{\rm max}$ is equivalent to the radius of the effective \textit{accreting surface}.

\subsection{Ionized envelope structure}\label{envelope}

\begin{figure}
    \centering
    \includegraphics[width=\columnwidth]{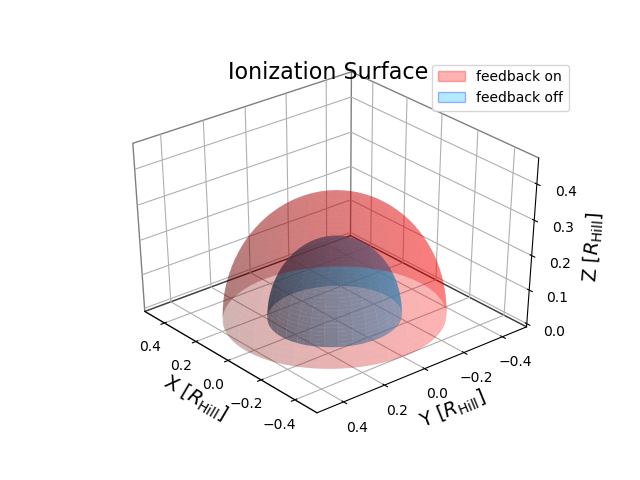} 
\caption{Ionization envelop (in Hill units) around the planet at the last output for models with and without feedback. The region is defined by the ionization parameter values $X \geq 0.5$.}
    \label{fig:envelope}
\end{figure}

In Figure \ref{fig:envelope}, we present a three-dimensional visualization of the ionized envelope plotted only by considering the ionization radius (corresponding to the maximum value obtained for the ionization radius). In the case with feedback mechanism activated, the ionized envelope is enhanced as shown in Section \S \ref{ionization_radius}.

\begin{figure}
    \centering
    \includegraphics[width=0.8\columnwidth]{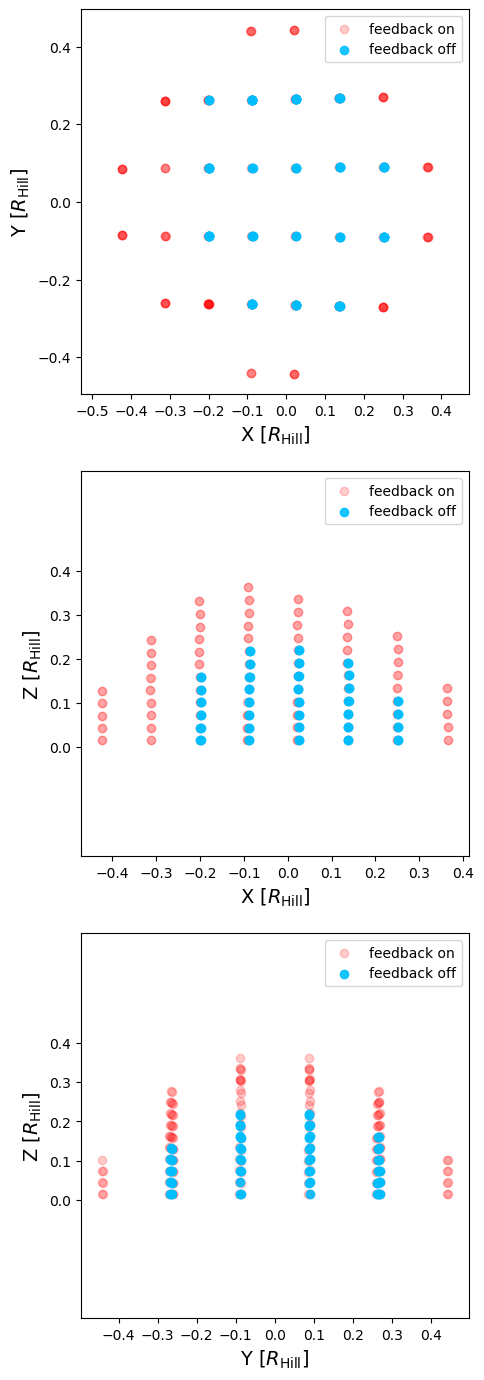} 
\caption{Different cuts of the ionization surface centered into the planet, comparing models with and without feedback. The top panel corresponds to the X-Y plane ($Z=0$), the middle panel to the Z-X plane ($Y=r_p$), and the bottom panel to the Z-Y plane ($X=0$). The geometry of the envelope is not perfectly spherical; it is obloid. The envelope shows a more significant enhancement in the Y direction, which corresponds to the radial direction.}
    \label{fig:ionization_surface}
\end{figure}

Upon closer inspection of Figure \ref{fig:envelope}, the envelope is not perfectly spherical, showing slight variations in its radius along different axes, emphasizing the complex nature of the accretion and ionization processes in play. For instance, the maximum radii along the Z (vertical), X (azimuthal), and Y (radial) axes are 0.337 $R_{\rm Hill}$, 0.364 $R_{\rm Hill}$, and 0.443 $R_{\rm Hill}$, respectively. In Figure \ref{fig:ionization_surface}, we present the geometry of the envelope as seen in different cut planes; X-Y ($Z=0$), Z-X ($Y=r_p$), and Z-Y ($X=0$), revealing that it is not perfectly spherical. As mentioned above, the ionization degree is more extensive in the Y (radial) direction. These dimensions are obtained from the final output of our simulation, but remain in quasi-steady state throughout the simulation after 300 planetary orbits. 

The anisotropic nature of the ionized envelope should be taken into account when calculating accretion rates and examining disk-planet interactions. For the scope of this paper, we simplify our model by considering only the maximum ionization radius, which we refer to as the \textit{ionization radius} $R_{\rm ion}$.

\subsection{Vertical velocity and radial inflow toward the planet}\label{vertical_velocity}

\begin{figure*}
\includegraphics[width=\textwidth]{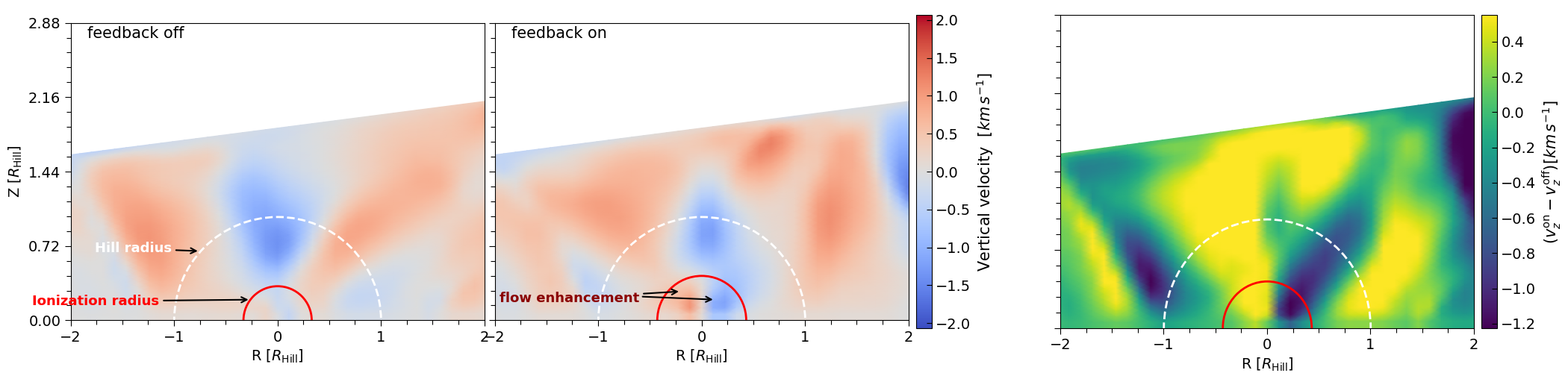}
\caption{Vertical velocity maps for a planet located at 
r=1, presented without (left panel) and with feedback (middle panel). Differences between the two states are shown in the right panel. Feedback amplifies both upward and downward flow near the planet. For r<1, feedback enhances upward velocity, as evidenced by higher positive values (yellow hues in the difference map $\sim 0.4$ (right panel). For r>1, feedback intensifies the downward flow of material, indicated by more negative values (blue hues in the difference map, right panel, $\sim -1.2$). These variations in vertical flow directionality signify feedback-induced alterations in material accretion and ejection dynamics.}
\label{fig:Vz}
\end{figure*}

The vertical velocity of the flow $v_z$ is affected by the planet's radiative feedback, which changes the dissipated energy associated with the turbulent viscosity. Consequently, the sound speed is also altered, and thus the region where shocks are expected to arise. This effect should be particularly noticeable in adiabatic models.

Figure \ref{fig:Vz} compares models with and without feedback of the vertical field velocity in the z-R plane for the last output (1000 orbits). We observe positive values of $v_z$ (reddish colors), indicating that the flow goes upward, and negative values (blueish) showing a falling flow into the midplane. Within the ionization radius and to the left of the planet's location (in stellocentric coordinates at $r \leq r_p = 1$), the flow is mainly upward. To the right of the planet (at $r > r_p = 1$), the flow moves inward. Once feedback is activated, the feedback mechanism significantly alters the flow dynamics, enhancing both the inward and upward flows. By computing the difference $v_z^{\rm on} - v_z^{\rm off}$ (shown in the rightmost panel of Figure \ref{fig:Vz}), we find that the inward component at $r \geq r_p =1$ is more enhanced than the upward flow. This is indicated by higher negative values in the third panel of Figure \ref{fig:Vz}, reaching -1.2 km/s at $r \geq r_p = 1$, compared to an enhancement of the upward flow to only +0.5 km/s (at $r \leq r_p = 1$). This results in an asymmetry between the inward and outward flows, with a net flow gain into the midplane, enhancing the planet's accretion rate when feedback is activated (see next Section \S \ref{sec:accretion}).

\begin{figure*}
\includegraphics[width=\textwidth]{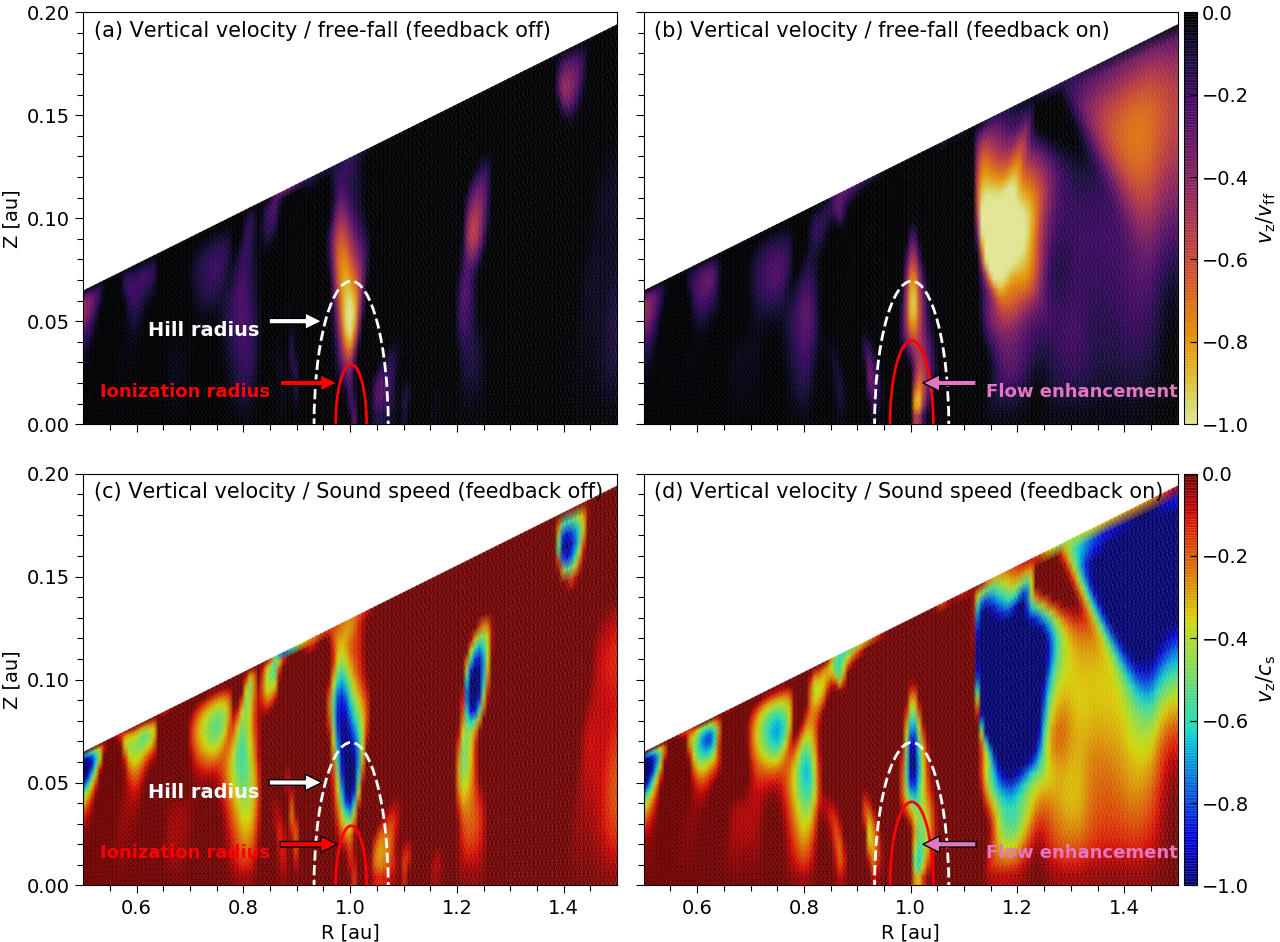}
\caption{Vertical velocity $v_{\rm z}$ of gas in comparison to free-fall and sound speed. Panel (a) displays the $v_{\rm z}$ velocity in the absence of planetary feedback. In this case, the flow comes to a halt ($v_{\rm z} \sim 0$) at the ionization radius directly above the planet's location, while the maximum vertical velocity ($v_{\rm z} \sim v_{\rm ff}$) is observed near the Hill radius (yellow region) in the same direction. In contrast, panel (b) shows the scenario with feedback, where the vertical flow is \textit{reduced} beyond the ionization radius but \textit{enhanced} within this radius (indicated by the pink arrow). Panel (c) displays the scenario without feedback, where the gas falls supersonically towards the planet from high altitudes beyond the ionization radius. Finally, panel (d) shows the scenario with feedback, where the supersonic falling gas is \textit{reduced} in these vertical regions beyond the ionization radius but notably \textit{enhanced} within this radius (indicated by the pink arrow). It is important to remark that in both scenarios with or without feedback, the falling gas stops precisely at the ionization radius in a direction directly above the planet's location.}
\label{fig:Vz_Cs}
\end{figure*}

Figure \ref{fig:Vz_Cs} shows the ratio of the vertical velocity field $v_{\rm z}$ compared with the free-fall gas velocity, i.e., $v_{\rm z}/v_{\rm ff}$ (top panels) where $v_{\rm ff} = \sqrt{2GM_{\rm p} / r}$, and the thermal sound speed, i.e., $v_{\rm z}/c_{\rm s}$ (bottom panels). In this case, the velocity field taken into account has the direction towards the planet; therefore, the figure shows negative values only.  

The top panels of Figure \ref{fig:Vz_Cs} show that the free-falling material is produced at high altitudes in a vertical column region above the planet's location. Also, the falling gas in these regions reaches supersonic velocities, i.e., $v_{\rm z} \sim v_{\rm ff} \sim c_{\rm s}$ (bottom panels of Figure \ref{fig:Vz_Cs}), which is consistent with other numerical simulations (e.g., \citealt{Tanigawa+2012, Szulagyi+2014}). This vertical velocity towards the planet attains its maximum value close to the Hill radius (white curve in Figure \ref{fig:Vz_Cs}). Noticeably, the deceleration of the gas is produced exactly at the ionization radius (red curve in Figure \ref{fig:Vz_Cs}). Furthermore, this deceleration at the ionization radius occurs with or without the radiative feedback. The ionization radius is, therefore, produced where the gas stops its free-fall journey toward the planet.

The fact that the gas deceleration occurs at the ionization radius is expected. And it can be understood as follows: ionization occurs when high temperatures are reached ($T \sim 10^{3}-10^{5} {\rm K}$). When the falling gas vertically approaches the planet's gravitational well, it piles up inside the circumplanetary envelope and starts to be compressed at some altitude due to high-pressure gradients. At this location, a temperature enhancement is produced when the kinetic energy of the falling gas is drastically reduced and converted into radiation and heat via viscous dissipation. Since the gas is optically thick, the cooling is inefficient, and a temperature increment is produced, boosting ionization.

Panels (a) and (b) of Figure \ref{fig:Vz_Cs} present a comparison of the vertical velocity of the gas in the absence and presence of the feedback, respectively. Upon activation of the feedback, the ionization radius expands, as discussed in Section \S \ref{ionization_radius}, resulting in a decrease in the free-fall region beyond the ionization radius at higher altitudes. However, \textit{within} the ionization radius, the radiative feedback exerts vertical forces on the gas, leading to a gas circularization that promotes the development of a \textit{planetary fallback rate} (pink arrows in Figure \ref{fig:Vz_Cs}). As a result, the net flux towards the planet is actually enhanced if the planet feedback is activated. The boundary of the fallback rate onset is the ionization radius.

\begin{figure*}
\includegraphics[width=\textwidth]{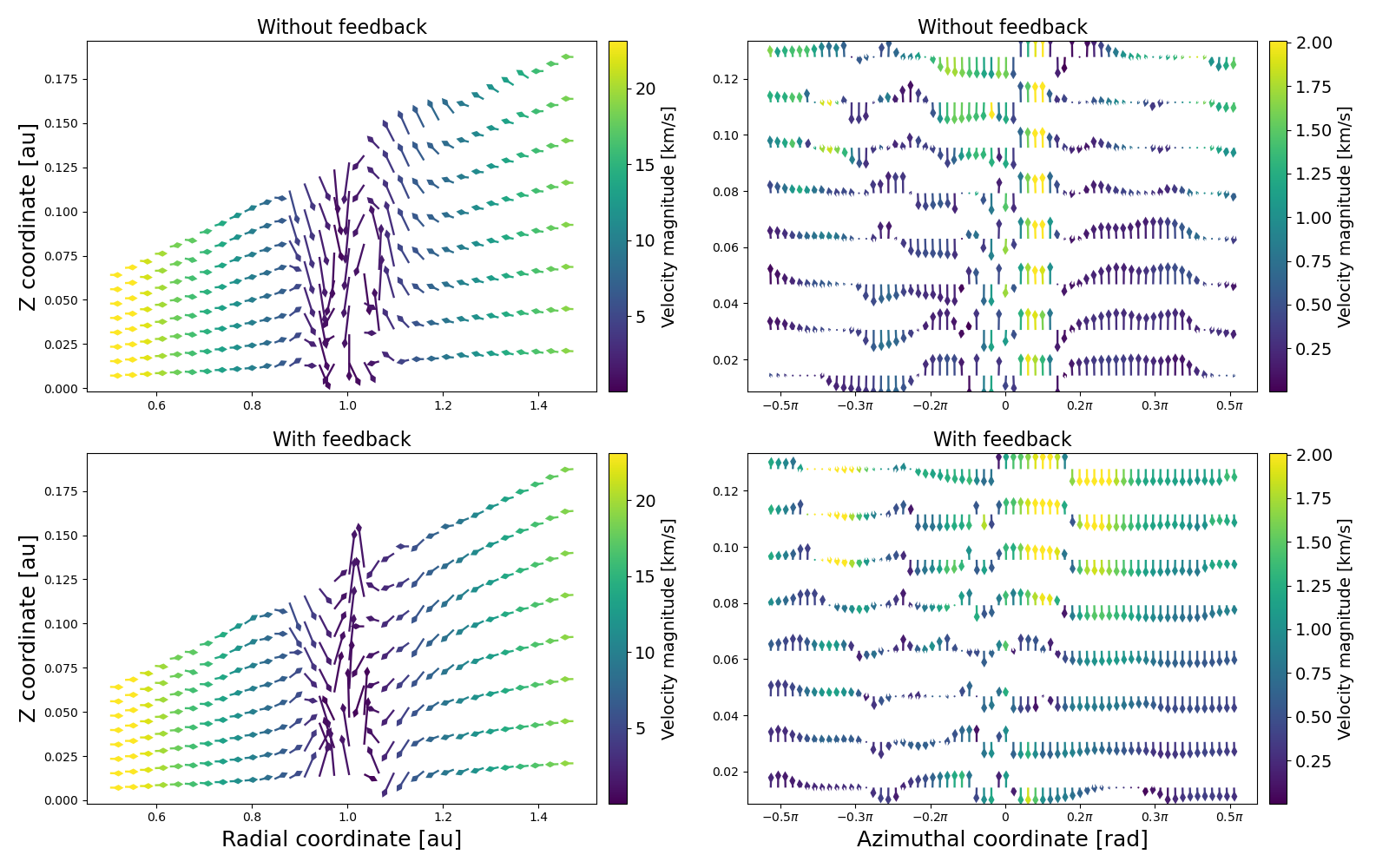}
\caption{Two-dimensional projections of the three-dimensional velocity field from simulations with and without feedback. The top left panel shows the velocity field projected onto the radial–Z plane for the no-feedback simulation, and the top right panel shows the corresponding projection onto the azimuthal–Z plane. The bottom panels display the same projections for the simulation with feedback. In all panels, arrows represent the normalized components of the 3D velocity vector after projection onto the respective plane, with arrow color denoting the projected velocity magnitude (in $km \, s^{-1}$). Arrow lengths are fixed to emphasize directional differences rather than absolute speeds, and a common color scale is used for direct comparison.}
\label{fig:Vz_vector}
\end{figure*}

Figure \ref{fig:Vz_vector} illustrates the directional behavior of the gas velocity. In the Z–R plane (left panels of Figure \ref{fig:Vz_vector}), and in the absence of feedback, the gas flows almost radially downward toward the planet. Once feedback is activated, however, the picture changes: in the immediate vertical vicinity of the planet, the downward flow is slowed or even reversed, with velocity vectors indicating upward or stalled motion. This outcome aligns with the expected effect of feedback, which tends to halt accretion—particularly under spherical accretion conditions. In a three-dimensional model, though, the gas begins to circularize around the planet as it is first driven upward over the planet and then redirected toward the equatorial plane. For instance, at larger radii, the material falls more rapidly than in the no-feedback case, as observed in the bottom-left panel of Figure \ref{fig:Vz_vector}.

In the Z–azimuth plane (right panels of Figure \ref{fig:Vz_vector}) the effect of feedback is also noticeably. When feedback is activated, the gas is redirected toward the equatorial plane, evidencing a change in flow direction. This behavior suggests that the radiative heating induces a convective-like circulation: gas is lifted above the planet due to increased local pressure and then turns laterally before eventually falling back toward the disk midplane and into the planet's envelope. Such a pattern highlights the role of feedback in driving the redistribution of angular momentum and mass, in agreement with the variations observed in the velocity maps presented in Figure \ref{fig:Vz}.

\begin{figure*}
    \centering
    \includegraphics[width=\textwidth]{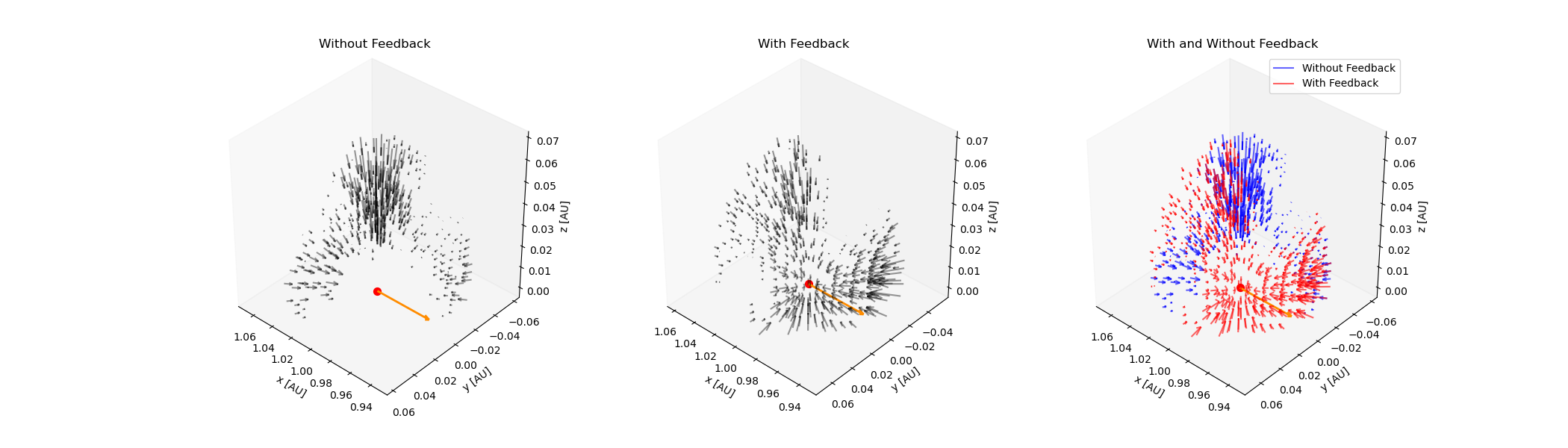} 
\caption{Radial velocity field of the gas towards the planet, plotted within the range of $0.5 R_{Hill}$ (approximately the ionization radius) to 1 $R_{Hill}$. The left panel shows the velocity field without feedback, and the middle panel shows the velocity field with feedback. The right panel overlays both simulations, with red representing the velocity field in the feedback case and blue in the no-feedback case. The orange arrow indicates the direction towards the star. When feedback is activated, there is an increase in the inflow towards the planet, particularly on the side closer to the star, highlighting the effect of feedback in enhancing accretion dynamics in this region.}
    \label{fig:radial_vel_out}
\end{figure*}

Beyond influencing the vertical flow structure, radiative feedback also modulates the radial inflow toward the planet. Figure~\ref{fig:radial_vel_out} presents the radial velocity field of gas directed toward the planet, focusing on the region between approximately $0.5,R_{\rm Hill}$ (roughly the ionization radius) and $1,R_{\rm Hill}$. The left panel shows the no-feedback case, where the inflow is predominantly vertical with a well-defined structure. In contrast, the middle panel—incorporating feedback—reveals an enhanced and more asymmetric inflow, particularly on the star-facing side of the planet, as indicated by the orange arrow marking the stellar direction. The right panel overlays the two simulations, with red representing the feedback case and blue the no-feedback case. This comparison demonstrates that radiative feedback amplifies the inflow in a spatially dependent manner, preferentially enhancing inflow on the star-facing side. Such asymmetry in the accretion flow suggests that ionization-driven feedback breaks the initial symmetry, thereby favoring a net inflow that may have important implications for the mass budget and structure of the circumplanetary disk.

\begin{figure*}
    \centering
    \includegraphics[width=\textwidth]{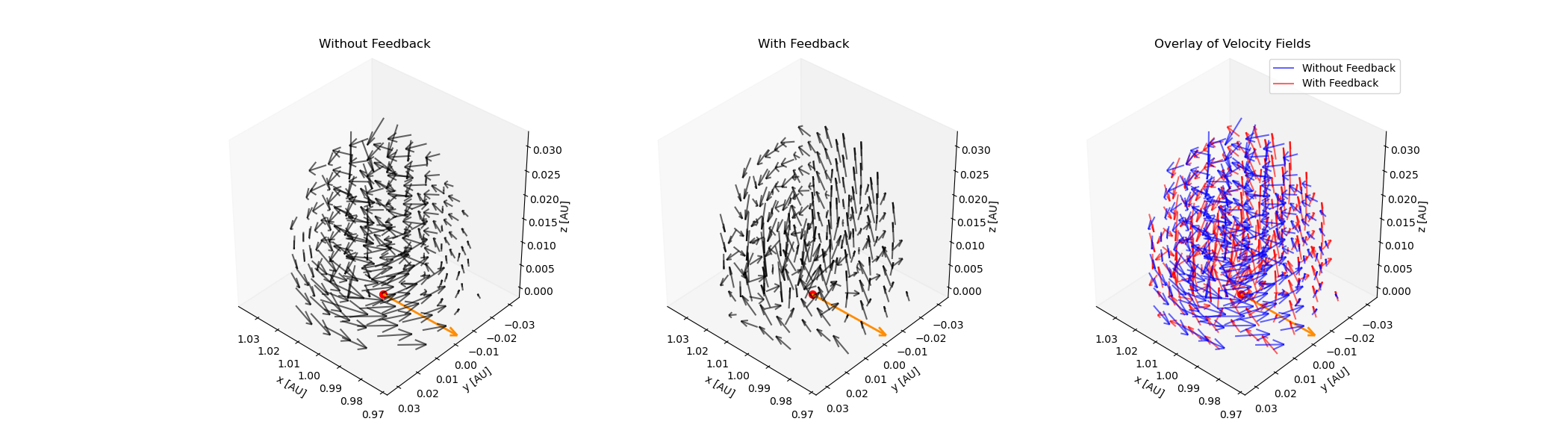} 
\caption{Velocity field of the gas within the region between 0 and 0.5 \( R_{\rm Hill} \), illustrating the dynamics of the bound envelope around the planet. The left panel shows the velocity field without feedback, where the gas exhibits a more ordered rotation, resembling a circumplanetary disk. The middle panel displays the velocity field with feedback, revealing a more turbulent motion, but with the gas still gravitationally bound to the planet. The right panel overlays both simulations, with red representing the feedback case and blue the no-feedback case, highlighting the structural differences introduced by feedback. The orange arrow in each panel points towards the central star.}
    \label{fig:radial_vel_in}
\end{figure*}

Figure \ref{fig:radial_vel_in} illustrates the velocity field within 0.5 $R_{\rm Hill}$, a region close to the ionization radius. In the absence of feedback, the gas displays a coherent rotational pattern, characteristic of a CPD. In contrast, when feedback is included, the gas motion becomes more turbulent, forming an envelope rather than a CPD yet still gravitationally bound to the planet. These findings agree with those of  \cite{Szulagyi+2016}, who report a hot circumplanetary envelope extending to a similar radial scale. Beyond 0.5 $R_{\rm Hill}$, the gas no longer rotates around the planet.

\subsection{Planet accretion rate and envelope luminosity}\label{sec:accretion}

\begin{figure*}
    \centering
    \includegraphics[width=\textwidth]{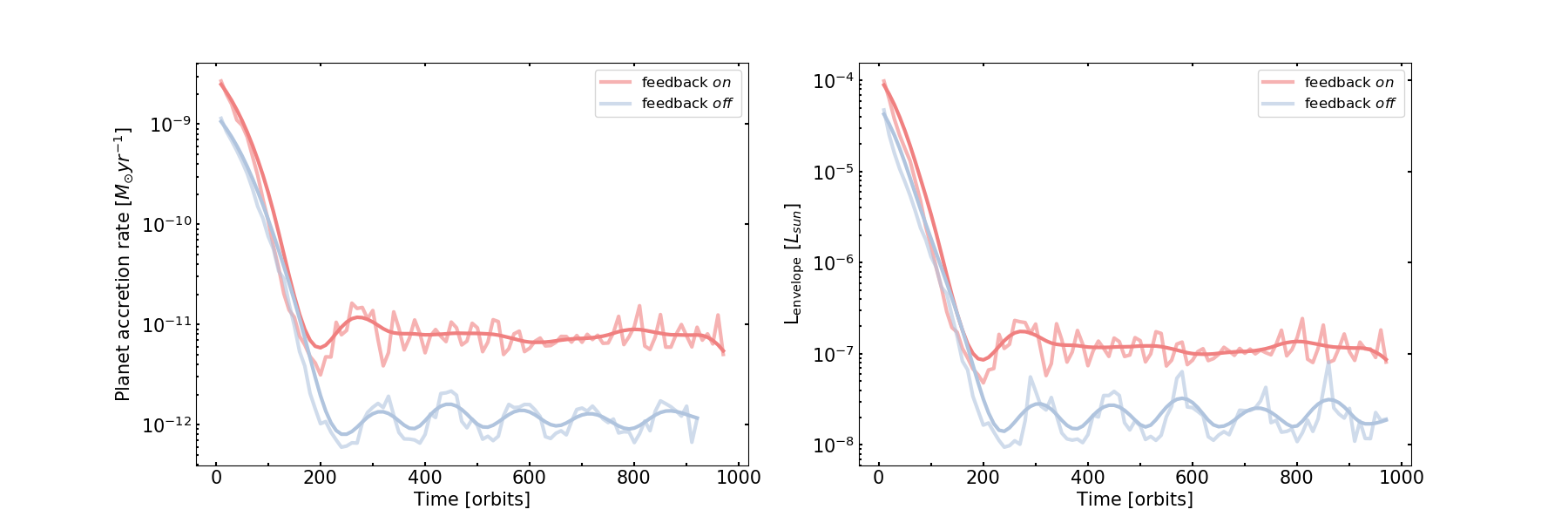} 
  
\caption{Left panel: evolution in time of the planet accretion rate (Equation \ref{eqMdot_final}) for two models with and without feedback. Once the disk reaches a quasi-stationary regime (after 300 planet orbits), the accretion rate is more or less constant. When the feedback is activated, the accretion rate in the quasi-steady regime is enhanced by a gain factor of about 5.4 (440\% enhancement). Right panel: evolution of the envelope luminosity obtained from Equation \ref{eq:Lshock}. Once the feedback is activated, the luminosity is enhanced by a similar factor as the accretion gain.}
    \label{fig:Mdot}
\end{figure*}

As discussed in Section \ref{PlanetAcretion}, the accretion rate towards the planet must be computed as the mass flux passing through the accreting surface. Figure \ref{fig:Mdot} (left panel) shows such flux as a function of time computed from Equation \ref{eqMdot_final}. In both scenarios—with and without feedback—the accretion rate stabilizes into a quasi-steady state after approximately 300 orbits. Specifically, the mean accretion rate during this stage is  $\langle \dot{M}_{\rm planet} \rangle \sim 1.2\times 10^{-12} M_{\sun} {\rm yr}^{-1}$ without feedback, and $\langle \dot{M}_{\rm planet} \rangle \sim 7.8\times 10^{-12} M_{\sun} {\rm yr}^{-1}$ with feedback. This corresponds to an accretion flow increase by a factor of $\Delta \dot{M}{\rm planet}/\dot{M}{\rm planet} = (\dot{M}{\rm planet}^{\rm on} - \dot{M}{\rm planet}^{\rm off})/\dot{M}_{\rm planet}^{\rm off} = 5.44$, or a 544\% enhancement relative to the no-feedback scenario. 

It is worth noting that due to resolution limitations, we cannot confirm that all the mass flux passing through the envelope will ultimately accrete onto the planet. However, given that the material resides within the gravitationally bound and Hill radius we can expect that material will feed the planet's envelope, and its luminosity will match to the accretion luminosity.

\begin{figure}
    \centering
    \includegraphics[width=\columnwidth]{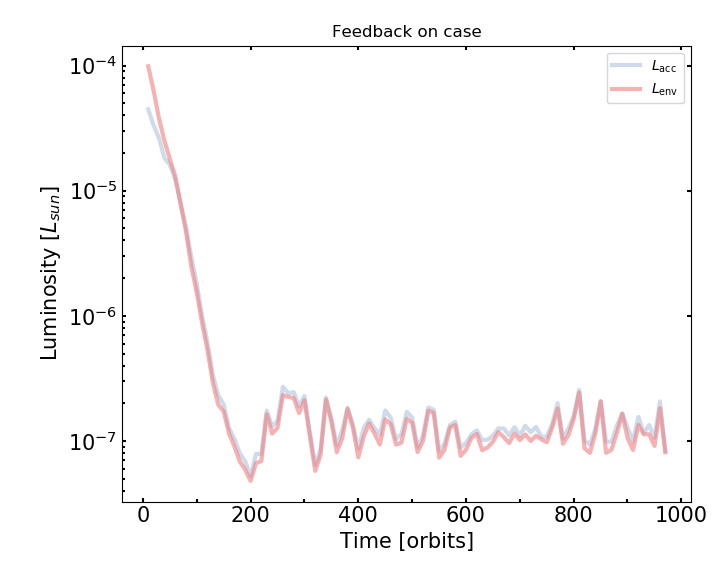} 
  \caption{Comparison of the bolometric luminosity of the envelope (as calculated by Eq. \ref{eq:Lshock}) and the accretion luminosity $L_{\rm acc}$ (calculated from the planet accretion), both for the model with feedback on. A great match is also observed for the feedback off model.}
    \label{fig:LaccLenv}
\end{figure}

The right panel of Figure \ref{fig:Mdot} shows the evolution of the envelope luminosity, as calculated from Equation \ref{eq:Lshock}. After the disk reaches a quasi-stationary state it stabilizes at a near constant value: $\langle L_{\rm envelope} \rangle \sim 2.25\times 10^{-8} L_\odot$ (feedback off), and $\langle L_{\rm envelope} \rangle \sim 1.4 \times 10^{-7}  L_\odot$ (feedback on). When feedback is \textit{turned on}, the envelope luminosity also experiences a gain factor of 5.4 (540 \% enhancement relative to the no-feedback case), exactly the same gain factor as the accretion rate, i.e., both $L_{\rm envelope}$ and $L_{\rm acc}$ show a match (see Fig. \ref{fig:LaccLenv}). Note that the calculation of the envelope luminosity $L_{\rm envelope}$ was obtained through the energy content (Eq. \ref{eq:Lshock}) and not through the equation for the accretion luminosity (Eq. \ref{eq:Lacc}), i.e., $L_{\rm acc} \simeq G M_{\rm p} \dot{M}_{\rm p}/R_{\rm in}$.  See next subsection \ref{sec:PlanetAcretion}.

\subsection{Planet accretion and luminosity rates at different radii}\label{sec:PlanetAcretion}

The envelope luminosity $L_{\rm envelope}$ computed from Equation \ref{eq:Lshock} must not be confused with the intrinsic planet luminosity $L_{\rm p}$. The intrinsic planet luminosity $L_{\rm p}$ is used as a free parameter representing the feedback inputs. The obtained envelope luminosity is related to the slow Kelvin–Helmholtz contraction of the planet's envelope. In fact, if we calculate the accretion luminosity $L_{\rm acc}$ using Equation \ref{eq:Lacc} replacing $\dot M_{\text{core}}$ with our computations of the accretion rate (Figure \ref{fig:Mdot}), \( M_{\text{core}} \) with \( M_J \), \( R_{\text{in}} \) with the mid-cell radius closest to the planet ($R_{\text{in}} \sim R_{\rm Hill}/10$ according to our numerical resolution), and \( R_{\text{out}} \) with \( r_{\text{Hill}} \), we find a remarkable match between \( L_{\text{envelope}} \) and the standard accretion luminosity \( L_{\text{acc}} \) (see Figure \ref{fig:LaccLenv}).

\begin{figure*}
    \centering
    \includegraphics[width=\textwidth]{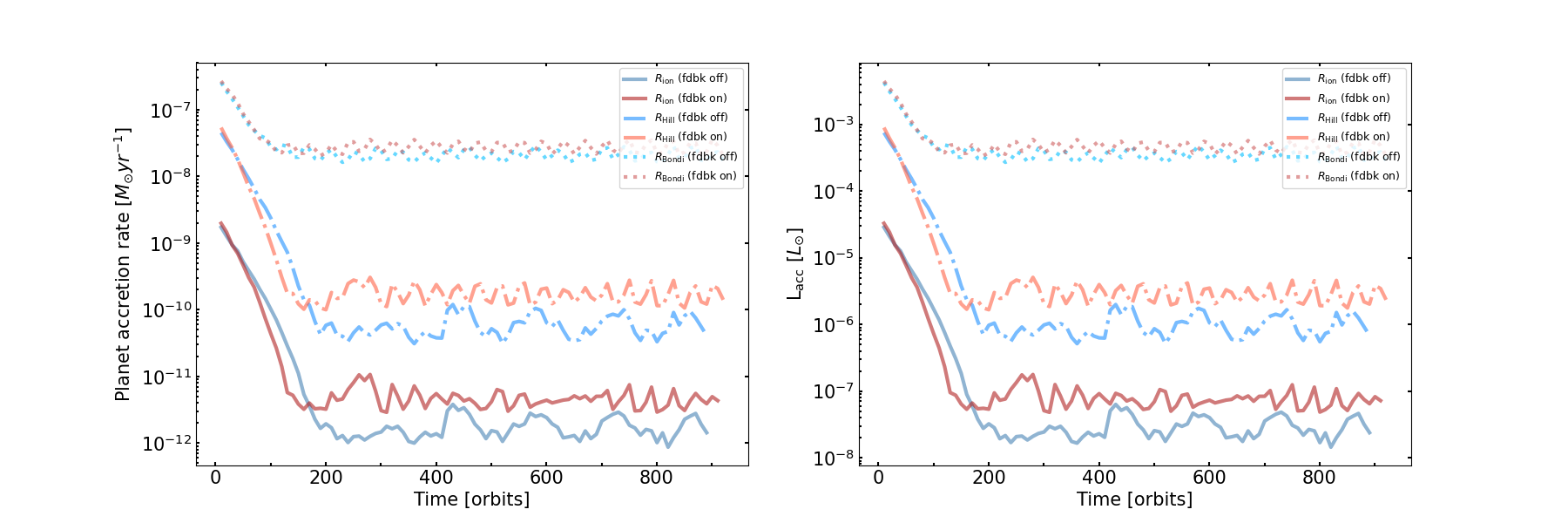} 
  
\caption{Left panel: planet accretion rate (Equation \ref{eqMdot_final}) for models with and without feedback computed at different radii: ionization radius $R_{\rm ion}$, Hill radius $R_{\rm Hill}$, and bound radius $R_{\rm Bondi}$. Right panel: evolution of the accretion luminosity $L_{\rm acc} = G M_{\text{p}} \dot M_{\text{p}} (1/R_{\text{in}} - 1/R_{\text{out}})$ for each radius (i.e., $R_{\rm ion}$, $R_{\rm Hill}$, and $R_{\rm Bondi}$), where $R_{\text{in}} = R_{\rm Hill}/10$ (corresponding to our grid resolution) and $R_{\text{out}} = R_{\rm Hill}$.}
    \label{fig:Mdot_radius}
\end{figure*}

Figure \ref{fig:Mdot_radius} compares the planet's accretion rate, computed at different radii—namely, the ionization radius $R_{\rm ion}$, the Hill radius $R_{\rm Hill}$, and the Bondi radius $R_{\rm Bondi}$—with its associated accretion luminosity $L_{\rm acc}$ (Eq. \ref{eq:Lacc}), obtained for a models with and without feedback. 
When the accretion rate is computed at a fixed radius, the accretion is always enhanced if the feedback is activated in the simulation, except when computed at the Bondi radius, where a small gain in accretion is observed.

In our setup, the different radii have the following values: $R_{\rm ion} \approx 0.3 R_{\rm Hill}$ and $R_{\rm Bondi} \approx 2.8 R_{\rm Hill}$. For instance, the accretion flow at $R_{\rm Hill}$ reaches a mean value of about $\dot{M}(R_{\rm Hill}) \sim 4.8 \times 10^{-11} M_\odot yr^{-1}$ while at the ionization radius reads $\dot{M}(R_{\rm ion}) \sim 1.2 \times 10^{-12} M_\odot yr^{-1}$. These values suggest that only $\dot{M}(R_{\rm ion})/\dot{M}(R_{\rm Hill}) = 2.5 \%$ of the gas passing through the Hill sphere reaches the ionization radius. When the feedback is activated, the accretion ratio reads $\dot{M}(R_{\rm ion})/\dot{M}(R_{\rm Hill}) = 5 \%$. Both $\dot{M}(R_{\rm ion})$ and $\dot{M}(R_{\rm Hill})$ are enhanced in the presence of feedback, but the increase is more pronounced for $\dot{M}(R_{\rm ion})$, leading to the observed change in such ratio.

The mean accretion luminosity after 300 orbits reaches approximately $2 \times 10^{-8} L_\odot$ for $R_{\rm ion}$ and approximately $7.9 \times 10^{-7} L_\odot$ for $R_{\rm Hill}$. This implies a gain factor of about 40 when calculating the accretion rates using $R_{\rm Hill}$ instead of the ionization radius. Similar conclusions are obtained for the case with feedback activated.

\subsection{Planet accretion and luminosity rates for different opacities}

We find that reducing the opacity from the fiducial value of $\kappa = 1 \text{cm}^2 \text{g}^{-1}$ to $\kappa = 0.01 \text{cm}^2 \text{g}^{-1}$, yields results that are not significantly different. The ionization radius remains practically unchanged, and the enhancement in accretion rate when feedback is activated is 4.84 (compared to 5.4 in the fiducial model), with a corresponding luminosity enhancement of 4.38. This indicates that the reduction in opacity does not substantially affect the overall accretion and luminosity outcomes.

\begin{figure*}[]
    \centering
\begin{minipage}[b]{0.45\linewidth}
    \includegraphics[width=\linewidth]{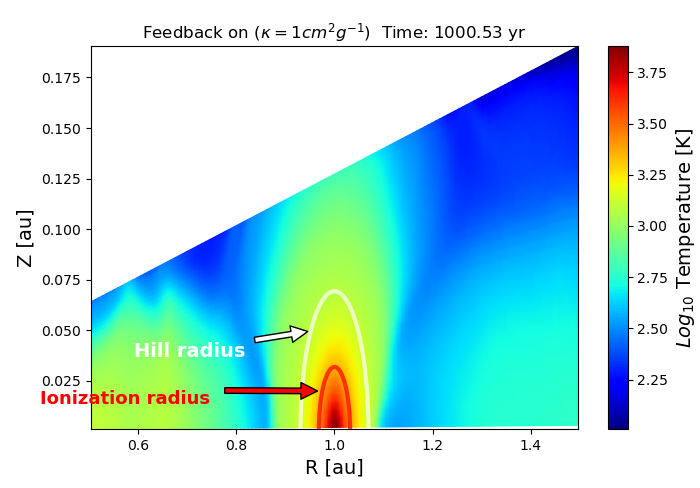}
\end{minipage}
\hspace{0.05\linewidth}
\begin{minipage}[b]{0.45\linewidth}
    \includegraphics[width=\linewidth]{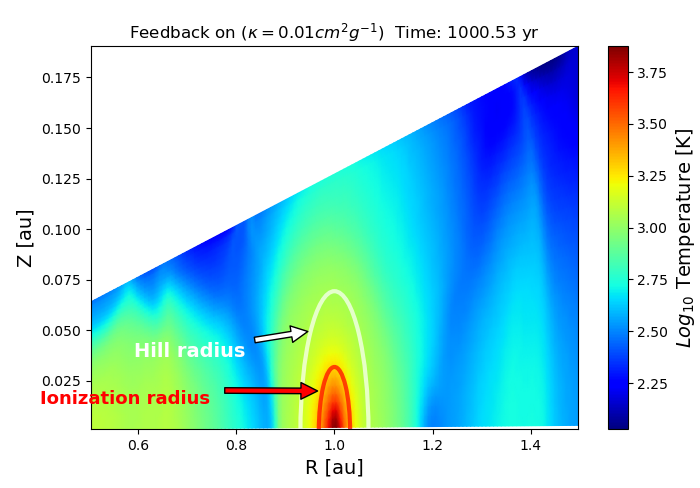}
\end{minipage}
    \caption{Temperature profiles in the z-plane, illustrating the thermal structure of the disk. Both sets of figures correspond to the case with feedback activated ($\kappa = 1 \, \text{cm}^2/\text{g}$ and $\kappa = 0.01 \, \text{cm}^2/\text{g}$).}
    \label{fig:temperature}
\end{figure*}

This behavior can be explained by the fact that the disk remains optically thick, particularly within the Hill radius, where high temperatures dominate the thermal and radiative processes (Figure~\ref{fig:temperature}). In optically thick regions, the temperature gradient adjusts to carry the necessary radiative flux, making the temperature structure relatively insensitive to the exact value of the opacity \citep{Hubeny1990}. As a result, the ionization levels remain nearly the same regardless of the change in opacity. The temperature structure of the disk shows minimal variation between the two opacity models (see Figure~\ref{fig:temperature}), further confirming that the disk's thermal properties are governed primarily by the optical depth rather than the absolute opacity value.

Therefore, the key results regarding accretion rates and the ionization radius are robust across different opacity values. Our findings demonstrate that, even with a reduced opacity, the physical processes dominating the circumplanetary environment remain unchanged, ensuring the reliability of the results presented.

\section{Discussion}\label{Discussion}

We have conducted 3D numerical simulations to track the evolution of a protoplanet embedded in a viscous disk, where the planet incorporates an additional radiative feedback term. We focused on thermal ionization, gas kinematics, planet accretion, and the connection between ionization and H-alpha emission. Our simulations reveal how the feedback influences the structure of the protoplanetary disk and the accretion process, providing insights into the formation and evolution of gas giants. In the following subsections, we discuss the key findings of our work and their implications in the context of planet formation and observational studies.

\subsection{Hot circumplanetary envelope and ionization effects}

We compute the gas thermal ionization degree $X$ (ranging from 0 to 1) using Equation ~\ref{eqX}. The onset of ionization is defined by the value $X=0.5$. The gas is considered neutral when $X<0.5$, while $X\geq0.5$ corresponds to partial to fully ionized gas. Utilizing this definition, we determined the ionization radius as the distance from the planet where the ionization degree is exactly $X=0.5$. 

We found that the protoplanetary disk is primarily neutral, except in a hot envelope (with temperatures up to $\sim 7,000$K) defined by a specific ionization radius close to the planet. The velocity field of the envelop is displayed in Figure \ref{fig:radial_vel_in}. In all our models, $R_{\rm ion}$ is smaller than the Hill radius, showing $R_{\rm ion}=\epsilon R_{\rm Hill}$, with $\epsilon \approx 0.2$ (feedback off) and $\epsilon \approx 0.4$ (feedback on), see subsection \S \ref{envelope}.

This envelope is consistent with the hot circumplanetary envelope with a radius of $\sim0.1-0.5R_{\rm Hill}$ reported by \cite{Szulagyi+2016}. The envelope's geometry depends on the inner content of energy and its radiative-convective transport. Thus, a non-isothermal model and the inclusion of radiative feedback play an important role in forming such a structure. 

\subsection{Local thermodynamic equilibrium}
In our post-processing calculation of the thermal ionization fraction via the Saha equation (\ref{eqX}), we assume local thermodynamic equilibrium (LTE). This assumption is well justified in the high-density midplane regions of protoplanetary disks (e.g., $10^{-8}$ to $10^{-11}$ g cm$^{-3}$) and at high temperatures ($\sim 5500$ K). Under such conditions, collisional timescales are significantly shorter than radiative timescales, ensuring that level populations remain in equilibrium with the local temperature (see for instance \cite{Mihalas+1984}).

\begin{figure}
    \centering
    \includegraphics[width=\columnwidth]{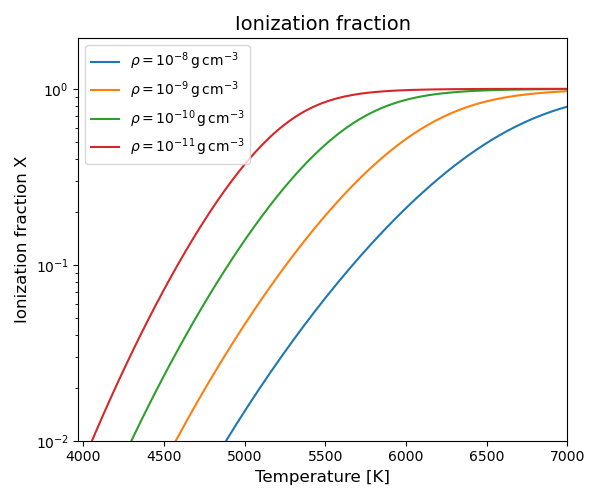} 
\caption{Ionization fraction X as a function of temperature for four representative midplane densities in a protoplanetary disk, ranging from $10^{-8}$ to $10^{-11}$ $g \, cm^{-3}$. The temperature range ($4000–7000 K$) and density values are typical of such environments. The curves, computed using the Saha equation (Eq. \ref{eqX}), indicate that lower densities result in higher ionization. For sufficiently high temperatures, the ionization fraction becomes independent of density.}
    \label{fig:ionization}
\end{figure}

To illustrate the dependence of the ionization fraction on both density and temperature, we compute curves using Equation \ref{eqX}. Figure \ref{fig:ionization} shows the ionization fraction as a function of temperature for representative densities ranging from $10^{-8}$ to $10^{-11}$ g cm$^{-3}$. In our model, the volume density is approximately $\rho \sim \Sigma / H \sim 10^{-10}$ g cm$^{-3}$, where $\Sigma$ and $H$ denote the surface density and scale height of the disk, respectively. Within the ionization radius, the gas reaches temperatures exceeding $\sim 5500$ K (Figure \ref{fig:temperature}). At this temperature threshold, the ionization fraction approaches $X \sim 0.5$, marking the transition from predominantly neutral to significantly ionized gas within the planetary envelope.

\subsection{Gas kinematics with planetary feedback}

By tracing the vertical velocity field of the gas, we observed that a vertical gas column falls towards the planet from high altitudes with a supersonic velocity. Once the gas reaches the ionization radius, its velocity rapidly decreases and stops. Regardless of whether the feedback is activated, the vertical flow always stops at the ionization radius, meaning that the ionization radius characterizes the stopping radius. 

When feedback is activated, the extent of the accreting column decreases(see Figures \ref{fig:Vz} and \ref{fig:Vz_Cs}). At first glance, this appears to contribute to a reduction in the accretion flow toward the planet as reported by \cite{Lambrechts+2017}. However, within the ionization radius, the planet's radiative feedback redistributes radiation and thermal energy within the planetary envelope, leading to gas recirculation (as indicated by the arrows in Figure \ref{fig:Vz}) enhancing the net flux mass trough the ionisation surface. We also notice that most of this mass flux is an unbound flow through the envelope. We explain this behaviour as follow:

In regions to the left of the planet, closer to the star at $r > r_p$, the downward flow of gas is particularly enhanced (while the upward flow is reduced) due to this recirculation. This promotes the development of a fallback rate, thereby increasing the net accretion toward the planet (see Figures \ref{fig:Vz} and \ref{fig:Vz_Cs}, where arrows indicate flow enhancement). An increase in the accretion rate translates into an enhancement of the envelope luminosity. This finding is consistent with two-dimensional numerical simulations of embedded radiative planets, where feedback stimulates stellocentric accretion \citep{Montesinos+2015}. 

This accretion enhancement is quite different from what \cite{Lambrechts+2017} found. The discrepancy may arise from the distinct nature of the modeled system (5-15 Earth mass cores vs Jupiter mass) and differing opacities. Additionally, they conducted their simulations over a shorter period (only 50 orbits, compared to 1000 in ours). Despite this, however, in our simulations the accretion rate appears to decrease during the first 200 orbits when the feedback mechanism is activated (see Figure \ref{fig:Mdot_radius}). The exploration of this discrepancy will be addressed in a future work but is outside the scope of this current study.

\subsection{Predicting shock zones and H-alpha emissions near the ionization radius}

The region just before the ionization radius, where the gas still falls at free-fall speed, may be considered the \textit{pre-shock} surface. At this location, the kinetic energy of the gas is still advected with it, making the dissipation of radiation inefficient, resulting in negligible thermal energy in the pre-shock region. However, once the gas reaches the ionization radius, its velocity drops significantly, and shocks should be produced at this frontier, converting part of the mechanical energy into radiation energy. Due to the limitations of our computational domain, it is not possible to detect or analyze this shock at the ionization frontier in our simulations; however, its occurrence is expected.

The relatively high ionization levels ($X\sim0.5$) in the post-shock region may also lead to H-alpha emission, which is a commonly used tracer of shock-heated gas in astrophysical systems and is expected to be generated by accreting sources such as protoplanets \citep[e.g.,][]{Aoyama2018, Aoyama+2020}. This highlights the importance of accurately determining the temperature, ionization level, velocity profile, and other variables needed to compute atomic lines. 

\subsection{Planetary accretion rate, luminosity and H$\alpha$ emissions}

We have shown that the ionization radius characterizes the stopping radius, delineating an ionized envelope around the planet. At this location, the gas velocity directed toward the planet reaches its apex before deceleration. The luminosity within this envelope, computed from the total energy density via $L_{\rm envelope} = d/dt \int_{X >= 0.5} \epsilon dV$ (Equation \ref{eq:Lshock}), matches good enough with the accretion luminosity, as defined by $L_{\rm acc} = \dot{M}(R_{\rm ion}) (G M_{\rm p}(1/R_{\rm in} - 1/R_{\rm out}))$ (Fig. \ref{fig:LaccLenv}). This agreement is expected; it ensues from the cooling of the planetary envelope via the thermal Kelvin-Helmholtz mechanism, captured by Equation \ref{eq:Lshock}, thereby verifying the consistency between thermodynamics and flow dynamics in our calculations.
Note that the accretion or envelope luminosity reported here is computed from the internal energy dissipation within an optically thick medium. It represents the energy released by Kelvin–Helmholtz contraction of the envelope rather than the effective luminosity measured directly at $\tau = 1$, which would be the observed luminosity of a forming planet.

The accretion rate at the Hill sphere differs from that at the ionization radius. Specifically, $\dot{M}(R_{\rm ion})/\dot{M}(R_{\rm Hill})$ equals 2\% in the feedback-off scenario and 5\% when feedback is on. Hence, a mere 2-5\% of the mass flow passing through the Hill sphere will eventually settle within the ionized envelope or on the planetary surface, which is consistent with the findings obtained by \cite{Marleau+2023}. The luminosity emanating from the Hill sphere surpasses that from the ionization radius (see Figure 1), implying that associating Hill luminosity with ionization-zone luminosity results in overestimation. It is thus important to correlate a characteristic luminosity from the ionized envelope with its corresponding accretion rate ($L_{\rm envelope} - \dot{M}(R_{\rm ion})$) for accurate estimation of H$\alpha$ luminosity, i.e., $L_{H \alpha} \propto \dot{M}(R_{\rm ion})$, which occurs solely in ionized media (and not at the Hill radius).

This proclivity to overestimate $L_{\rm acc}$ by utilizing $\dot{M}(R_{\rm Hill})$ may account for the elusive H$\alpha$ emissions from accreting planets in observational data (e.g., \citealt{Zurlo+2020}), contrary to theoretical models predicting sufficiently elevated accretion rates to stimulate such emissions (e.g., \citealt{Aoyama2019}). Our findings reveal that $\dot{M}(R_{\rm ion})$ and $\dot{M}(R_{\rm Hill})$ vary by nearly two orders of magnitude (same variation in the accretion luminosities (refer to Figure \ref{fig:Mdot_radius})). Also, our results suggest that the presence of ionized envelopes around accreting protoplanets at 1 AU does not necessarily lead to direct observational signatures in H$\alpha$ due to the extremely high optical depths. This suggests that triggered H$\alpha$ emissions should likewise be revised downward, which is particularly important when analyzing non-detection statistics.

\section{Conclusions}

We conduct 3D numerical simulations to follow the evolution of Jupiter-like planets embedded in a viscous disk, incorporating a non-stationary energy equation to treat gas and radiation energy transport. We also include an extra term to account for the radiative feedback of planets. We focused on the vicinity of planets, identifying an ionized envelope surrounding them. This envelope is defined by an ionization radius, which also characterizes the stopping radius of the falling gas toward the planet. Examining the vertical velocity field shows that the gas falls toward the planet at supersonic speeds. Upon reaching the ionization radius, the gas experiences a dramatic decrease in velocity. At this point, part the kinetic energy of the gas is converted into radiation and heat.

The inclusion of the radiative feedback enhances gas accretion toward the planet. The gas flow within the ionization radius lies on different 3-D trajectories, but particularly, it moves both upward and downward in the vertical direction. In the presence of feedback, both directions are boosted; however, the enhancement is more pronounced in the downward direction in regions further out from the planet's position at radii $r \geq r_p$ (regions on the right side of the planet, assuming the star is located to the left). As a result, the net flux toward the planet increases. Also, because the accretion rate is higher when the feedback is switched on, the liberated energy $L_{\rm acc}$ within the ionization radius also increases, leading to an expansion of the ionization radius. In our models, the ionization radius is considerably smaller than the Hill radius; specifically, $R_{\rm ion}$ is equal to 0.2 $R_{\rm Hill}$ (feedback off) and 0.4 $R_{\rm Hill}$ (feedback on).

We compare the accretion rate toward the planet at the Hill sphere and at the ionization radius, showing that $\dot{M}(R_{\rm ion})/\dot{M}(R_{\rm Hill})$ is equal to 2\% (when the feedback is off) and 5\% (in the feedback-on case). This ratio indicates that only a small fraction of the gas passing through the Hill sphere will end up inside the envelope (or reach the planet's surface). If H$\alpha$ emissions are produced due to the planet's accretion, they will only be triggered in an ionized medium. Since the ionization radius characterizes such a region, we propose computing the accretion rate at $R_{\rm ion}$, rather than at the Hill radius, to determine the luminosity associated with H$\alpha$, e.g., $L_{H \alpha} \propto \dot{M}(R_{\rm ion})$. Moreover, since $\dot{M}(R_{\rm ion}) \ll \dot{M}(R_{\rm Hill})$, the accretion luminosity characterized by $\dot{M}(R_{\rm ion})$ will be much smaller than $L_{\rm acc}$ at the Hill sphere; therefore, $L_{H \alpha}$ should also be much smaller if computed at $R_{\rm ion}$ rather than at $R_{\rm Hill}$ as is normally calculated. In this context, realistic computations of $L_{H \alpha}$ and corresponding accretion rates into the planet are probably smaller than those assumed through the Hill sphere. This may be important to consider when explaining the statistics of non-detections of H$\alpha$ emission as accretion tracers for forming planets. 

Finally, the study tested two different constant opacity values, $\kappa = 1$ and $\kappa = 0.01 , \text{cm}^2/\text{g}$, and found that both yield the same conclusions regarding accretion and luminosity outcomes. This is due to the fact that the disk remains optically thick in both cases, making the temperature structure and the physical processes largely insensitive to the specific opacity value.

We summarize our conclusions as follows:

\begin{itemize}
  \item Jupiter mass protoplanets should be surrounded by an ionized envelope with ionization radius $R_{\rm ion} \sim 0.2-0.4 R_{\rm Hill}$. Such distance defines the truncation radius of the envelope
  \item The ionization radius could be used to define the \textit{accretion radius} of the protoplanet envelope
  \item The 3D radiative feedback of the planet enhances the planet accretion rate $\dot{M}$ and, consequently, the envelope luminosity $L_{\rm envelope} = d/dt \int_{X \geqslant 0.5} \epsilon dV$. This luminosity is (as expected) the same accretion luminosity, i.e., $L_{\rm acc} \simeq G M_{\rm core} \dot{M}_{\rm core}/R_{\rm core}$ 
  \item Inside the ionized envelope, favorable conditions to generate Hydrogen emission lines are produced (e.g., H$\alpha$)
  \item Since H$\alpha$ emissions are triggered in an ionized medium only, the associated accretion rate $\dot{M}_{\rm H \alpha}$ should be linked to the accretion at the ionization radius $\dot{M}(R_{\rm ion})$ (rather than at the Hill radius). i.e., $L_{H\alpha} \propto \dot{M}(R_{\rm ion})$
  \item $\dot{M}(R_{\rm ion})/\dot{M}(R_{\rm Hill}) \sim 2 - 5\%$ (depending on the feedback). Since $\dot{M}(R_{\rm ion}) \ll \dot{M}(R_{\rm Hill})$, this suggests that accretion rates associated with H-alpha luminosity $\dot{M}_{\rm H \alpha}$ (and consequently $L_{\rm H\alpha}$) are significantly lower than those considered in the literature computed through the Hill sphere.
\end{itemize}


\begin{acknowledgments}
The authors thank the anonymous referee for their constructive feedback, which significantly improved the clarity of this work. The authors acknowledge support by ANID, -- Millennium Science Initiative Program -- NCN19\_171. This project has received funding from the European Research Council (ERC) under the European Union Horizon Europe programme (grant agreement No. 101042275, project Stellar-MADE). MM acknowledges financial support from FONDECYT Regular 1241818 and thanks Amelia Bayo and Johan Olofsson for insightful discussions during the early stages of this project. MS acknowledges support by Agencia Nacional de Investigaci\'on y Desarrollo (ANID) through FONDECYT postdoctoral 3210605. MRS acknowledges support from FONDECYT (grant 1221059). MPR is partially supported by PICT-2021-I-INVI-00161 from ANPCyT, Argentina. MPR and OMG are partially supported by PIP-2971 from CONICET and by PICT 2020-03316 from ANPCyT, Argentina. 
\end{acknowledgments}

%


\bibliography{astro}{}
\bibliographystyle{aasjournal}



\end{document}